\documentclass[aps,prb,showpacs,showkeys,twocolumn]{revtex4-1}
\usepackage[T2A,OT1]{fontenc}
\usepackage[cp1251]{inputenc}
\usepackage[russian,ukrainian,english]{babel}
\usepackage[dvips]{epsfig}            
\usepackage{latexsym}
\usepackage{amsfonts}
\usepackage{xcolor}[2007/01/21]
\usepackage[dvipdfm,bookmarks=true,bookmarksnumbered,bookmarksopen]{hyperref}
\hypersetup{pdfpagemode=FullScreen,colorlinks=true,
citecolor=green,unicode,bookmarksopenlevel=3,
pdftoolbar=true,pdfauthor={Helen Gomonay},pdftitle={Spin transfer in AFM}}
\usepackage{amsbsy}         

\begin{document}

\title{Symmetry and the macroscopic dynamics of antiferromagnetic materials in the presence of spin-polarized current}
\author{Helen V. Gomonay}\author{Roman V. Kunitsyn}
\affiliation {
 National Technical University of Ukraine ``KPI''\\ ave Peremogy, 37, 03056, Kyiv,
Ukraine}

\author{Vadim M. Loktev}
\affiliation {Bogolyubov Institute for Theoretical Physics NAS of
Ukraine,\\ Metrologichna str. 14-b, 03143, Kyiv, Ukraine}

\begin{abstract}
Antiferromagnetic (AFM) materials with zero or vanishingly small
macroscopic magnetization are nowadays the constituent elements of
spintronic devices. However, possibility to use them as active
elements that show nontrivial and controllable magnetic dynamics
is still discussible. In the present paper we extend the
phenomenologic theory [A.F.Andreev, V.I.Marchenko, Sov. Phys. ---
Uspekhi, 23 (1980), 21] of macroscopic dynamics in AFMs for the
cases typical for spin-valve devices. In particular, we consider
the solid-like magnetic dynamics of AFMs with strong exchange
coupling in the presence of spin-polarized current and give the
general expression for the
current-induced  Rayleigh dissipation function in terms of the rotation vector for different types 
of AFMs. Basing on the analysis of linearized equations of motion
we predict the current-induced spin-reorientation and AFM
resonance, and found the values of critical currents in terms of
AFMR frequencies and damping constants. The possibility of
current-induced spin-diode effect and second-harmonic generation
in AFM layer is also shown.

\end{abstract}
\date{\today}
\maketitle
\section{Introduction}
\label{sec_Intro} Spin polarized electric current flowing through the
nanoscale magnetic multilayers  can exert torque on a ferromagnetic
(FM) layer. This effect provides a potentially useful method for
magnetization switching and is widely used in spintronics (see,
e.g., review \cite{stiles-2008-320} and references therein). The
phenomenon of spin transferred torque (STT) was first suggested by
Berger \cite{Berger:1996} and Slonczewski \cite{SLonczewski:1989,
  Slonczewski:1996} in their seminal papers,
based on a rather general argument: conservation of the angular/spin
momentum for a system that includes both free (itinerant) electrons
and localized moments. Well-known
 expression for the Slonczewski's STT term,  $\mathbf{T}_{\mathrm{STT}}\propto J[\mathbf{M}\times[\mathbf{M}\times\mathbf{p}_{\rm
cur}]]$ (where $J$ is an electrical current and $\mathbf{p}_{\rm
cur}$ is the current polarization, notation $\times$ means
cross-product) is treated straightforward from the microscopic
considerations for a FM with magnetization vector $\mathbf{M}$. On
the other hand, there are many systems with the complicated
magnetic structure and vanishingly small (or zero) magnetization
(like antiferromagnets (AFMs) or spin glasses). So, the question:
``Can we observe STT phenomena in these materials and how should
we describe their dynamics in the presence of spin-polarized
current?'' has naturally appeared soon after the reminded papers
\cite{Berger:1996, SLonczewski:1989,
  Slonczewski:1996}. Recent experiments \cite{Tsoi:2007,
Urazhdin:2007, tang:122504, Dai08, Tsoi-2008} give an indirect
evidence that the spin-polarized current may also influence the
state of AFM layer. In addition, the experiments show that the
insertion of AFM layer drastically reduces the critical current
density of magnetization switching in the standard spin-valves
\cite{lee:08G517}. These phenomena, along with the effect of
anisotropic magnetoresistance \cite{Park:PhysRevB.62.1178,
Park:2010arXiv1011.3188P, Shick:2010} and high eigen frequencies
of the magnetic modes, make AFMs the promising candidates for the
use in spintronic devices.

The problem of STT in AFMs has two aspects. The first aspect,
considered theoretically by several groups, \cite{xu:226602,
duine:2007, Haney:2007(2), Tserkovnyak:PhysRevLett.106.107206}
concerns the \emph{ability of AFM materials} to polarize the
electric current that flows through it. The second, --
``inverse''-- aspect, related to the above mentioned experiments,
concerns the \emph{ability of spin-polarized current} to influence
the state of AFM layer. In the present paper we are concentrated
on the second aspect and study the dynamics of AFM layer coupled
electrically with the fixed FM layer (polarizer). Our main goal is
to develop a general formalism able to describe the magnetic
dynamics of AFM layer induced by already polarized spin current.

At present, theoretical models for interpretation of STT phenomena
in AFMs are based on microscopic calculations for some model
systems, \cite{xu:226602, duine:2007,
Haney:2007(2),Vedyayev:2008JAP...103gA721M} on the Landau-Lifshitz
equations for magnetic sublattices \cite{gomo:2008E, gomo:2010}
added with the Slonczewski's term, continuity equations for
two-sublattice AFMs \cite{Gulyaev:2010} or phenomenological
approach in the framework of nonequilibrium thermodynamics.
\cite{Tserkovnyak:PhysRevLett.106.107206} In most of these models
the magnetic structure of AFM is considered as a set of magnetic
sublattices or, in other words, embedded FM lattices that are
coupled by the exchange interaction. On the other hand, according
to Andreev and Marchenko \cite{Andreev:1980}, such an approach has
some shortcomings: \emph{i}) application of the Landau-Lifshitz
equations (added with the  STT term \cite{gomo:2008E,
gomo:2010,Tserkovnyak:PhysRevLett.106.107206}) for each of
magnetic sublattices is ``sometimes questionable though gives the
correct results in most cases'' \cite{Andreev:1980}; \emph{ii})
description of macroscopic (long-wave) dynamics in terms of
sublattice magnetizations is redundant for many AFMs (with three
and more sublattices).

To overcome an excessive detailing and model assumptions, Andreev
and Marchenko \cite{ Andreev:1980, Andreev:1976} have developed the
general phenomenological (``hydrodynamic-like'') theory for the
description of macroscopic magnetic dynamics of the materials with
strong exchange coupling between the sublattices. They have shown
that any magnetic structure (including multisublattice magnets) can
be characterized with the at most three mutually orthogonal magnetic
vectors that could be introduced from symmetry considerations only,
irrespective to microscopic mechanisms of AFM ordering.

In the present paper we make an attempt to extend such an approach
to the systems that are under the action of spin-polarized current.
Starting from the spin conservation principle we derive an
expression for the Rayleigh dissipation function and corresponding
dynamic equations in terms of macroscopic variables that describe
solid-like rotation of the magnetic moments\footnote{~We exclude any
possible solid-like rotation of the crystal lattice assuming that
the sample is fixed and consider rotation of the magnetic subsystem
only. However, STT can, in principle, give rise to rotation of the
crystal lattice and/or to the combined magneto-mechanical
oscillations, due to the spin-lattice coupling. These problems
though interesting are out of scope of this paper.} in the presence
of the external current-induced forces (torques). From general
considerations we show  that the effect of spin torque in AFMs can
be indeed as strong as in FM materials and analyze some typical
features of current-driving dynamics for AFMs with the collinear,
planar and nonplanar ordering.

\section{Dynamics of antiferromagnets within Lagrange formalism}\label{sec_Lagrange}
\begin{figure}[htbp]
  \includegraphics[width=1\columnwidth]{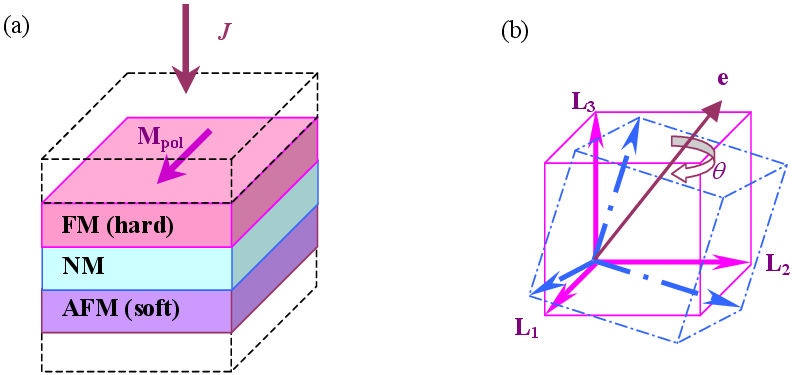}
  \caption{(Color online) \textbf{Effect of spin transfer torque on AFM layer}.
  (a)
Element of spin-valve structure consisting of FM (hard) and AFM
(soft) layers and NM spacer. The FM layer  polarizes the spin
current $\mathbf{J}$ along $\mathbf{M}_\mathrm{pol}$ direction. The
positive-valued current flows from FM to AFM layer. (b)
Macroscopic current-induced dynamics of an arbitrary AFM with
strong exchange coupling. Equilibrium magnetic structure is
characterized by the three mutually orthogonal magnetic vectors
$\mathbf{L}_1$, $\mathbf{L}_2$, and $\mathbf{L}_3$ (solid arrows).
Any deflection from equilibrium (dash arrows) can be viewed as a
solid-like rotation around the instantaneous axis $\mathbf{e}$
through the angle $\theta$.
 \label{fig_geometry}}
\end{figure}

Hereafter we consider the system (like a fragment of typical
spin-valve) which consists of the hard FM, nonmagnetic (NM), and AFM
layers (see Fig.~\ref{fig_geometry} a). FM layer works as a
polarizer for conduction electrons. We assume the FM vector
$\mathbf{M}_{\mathrm{pol}}$ (and, correspondingly, current
polarization
$\mathbf{p}_{\mathrm{curr}}\|\mathbf{M}_{\mathrm{pol}}$) to be
fixed and unchanged within FM layer. NM layer works as a buffer
which excludes the direct magnetic (exchange or dipole-dipole)
interaction between FM and AFM layers. Besides, the thickness of
NM layer is supposed to be below the spin-diffusion length, in
order to keep the extent of current polarization almost constant
throughout the layer. AFM layer is supposed to be rather thin, so,
that \emph{i}) it can be set into motion due to STT that takes
place in a thin region near NM/AFM interface and \emph{ii}) its
magnetic structure can be assumed homogeneous enough within the layer (so
called macrospin approximation). An electric current flows
perpendicular to the film plane in the case of conducting AFM (so
called CPP configuration, see Fig.~\ref{fig_geometry} a) and
between FM and AFM layers in the case of nonconducting AFM (CIP
configuration). AFM layer can be either compensated AFM, or weak
FM, or even spin glass. Following the terminology used for
spin-valve structures we will refer to AFM layer as a soft one.

\subsection{Comparison of spin transfer torques in ferro- and antiferromagnets}
To clarify the main ideas of the present paper we start from short
review of  STT phenomenon in ferromagnets where it manifests the
spin conservation principle for the system consisting of free
(itinerant) and localized spins. \cite{Slonczewski:JC1999L261}

Equations for spin transfer torque in a soft FM  can be obtained from
the balance equation for the magnetic moment:
\begin{eqnarray}\label{eq_dynamics_2}
   &&\frac{d\mathbf{M}_\mathrm{FM}}{dt}=\nabla\cdot
\hat{\boldsymbol{\Pi}},\nonumber\\ &&\textrm{or}\quad
\frac{dM_{\mathrm{FM}}^{(k)}}{dt}=\frac{\partial\Pi_{kl}}{\partial x_l}, 
\end{eqnarray}
where $\hat{\boldsymbol{\Pi}}$ is the 2-nd rank tensor of the
magnetization flux density induced, in particular, by
spin-polarized current. The l.h.s. of Eq.~(\ref{eq_dynamics_2})
includes the torque $\mathbf{T}_{\mathrm{FM}}$ produced by the
effective magnetic field and internal damping (not to be interchanged with the external spin transfer torque). Thus, in the
absence of a current the Eq.~(\ref{eq_dynamics_2}) takes a form of a
standard Landau-Lifshits-Gilbert equation:
\begin{eqnarray}\label{eq_LLG}
 \frac{d\mathbf{M}_\mathrm{FM}}{dt}&\equiv&\frac{\partial \mathbf{M}_\mathrm{FM}}{\partial
  t}\\
& +&\underbrace{\gamma
  \mathbf{H}_{\mathrm{eff}}\times\mathbf{M_\mathrm{FM}}+\frac{\alpha_{\mathrm{FM}}}{M_\mathrm{FM}}\frac{\partial \mathbf{M}_\mathrm{FM}}{\partial
  t}\times\mathbf{M}_\mathrm{FM}}_{\mathbf{T}_{\mathrm{FM}}}=0,\nonumber
\end{eqnarray}
where $\gamma$ is the gyromagnetic ratio\footnote{~In many
textbooks and papers the modulus of gyromagnetic ratio is used,
this gives rise to an opposite sign before the second term in
(\ref{eq_LLG}}, $\mathbf{H}_{\mathrm{eff}}$ is an effective
magnetic field, $\alpha_{\mathrm{FM}}$ is the Gilbert damping
parameter.

In the macrospin approximation (thin, magnetically homogeneous soft FM
layer) the r.h.s. of Eq.~(\ref{eq_dynamics_2}) can be expressed in
terms of the magnetization flux (per unit square) transferred by
the current $J\equiv jS_{\mathrm{FM}}$ that flows through the FM
layer perpendicularly to its surface:
\begin{equation}\label{eq_flux_1}
\nabla\cdot \hat{\boldsymbol{\Pi}}=\frac{\hbar\gamma
S_{\mathrm{FM}}}{|e|v_{\mathrm{FM}}}j\hat\epsilon\mathbf{p}_{\mathrm{curr}},
\end{equation}
where $S_{\mathrm{FM}}$ and $v_{\mathrm{FM}}$ are the surface area and the volume of
the FM layer, respectively, $\hbar$ is the Plank constant, $e$ is an electron
charge, and, as above, $\mathbf{p}_{\mathrm{curr}}$ is the
direction of current polarization,
$|\mathbf{p}_{\mathrm{curr}}|=1$. Second-rank tensor
$\hat\epsilon$ is proportional to the phenomenological (material) constant $\epsilon_{\mathrm{s-f}}$ that
describes the efficiency of spin-flip scattering of the carriers
at the NM/FM interface and depends upon  the details of
$sd$-exchange interaction between the free and localized
electrons. The structure of the tensor in general depends upon the
magnetic symmetry of the soft FM layer. In the particular case of
 isotropic layer\cite{Slonczewski:1996,Kohno:2007} (only the exchange symmetry is taken into
 account)
\begin{equation}\label{eq_Slon_1}
  \hat\epsilon =\epsilon_{\mathrm{s-f}}\left(\hat 1-\mathbf{e}_{M}\otimes
  \mathbf{e}_{M}\right),\quad \mathbf{e}_{M}\equiv
  \frac{\mathbf{M}_{\mathrm{FM}}}{M_{\mathrm{FM}}},
\end{equation}
where the symbol $\otimes$ means the direct tensor product, $\hat 1$ is a unit matrix.

In contrast to ferromagnets, AFMs have complicated,
multisublattice magnetic structure and vanishingly small or zero
equilibrium macroscopic magnetization. However, the set of dynamic
equations for AFMs includes the same spin-flux balance equation
(\ref{eq_dynamics_2}) as for FMs. The only difference is that in the
case of AFMs the vector $\mathbf{M}_{\mathrm{AFM}}$ of macroscopic magnetization
is induced by anisotropic Dzyaloshinskii-Moria exchange
interactions, by the external magnetic field and, what is essential, by the
\emph{solid-like rotation} of the magnetic lattice, as was shown
by Bar'yakhtar and Ivanov \cite{Bar-june:1979E} by the direct
analysis of the Landau-Lifshitz equations and in
Ref.~\onlinecite{Andreev:1980} basing on Noether's theorem. For
example, in the collinear compensated AFM with an AFM vector
$\mathbf{L}$ the macroscopic magnetization $\mathbf{M}_{\mathrm{AFM}}\propto
[\mathbf{L}\times \dot{\mathbf{L}}]$.

It follows from the balance equation (\ref{eq_dynamics_2}) that
the spin-flux transferred by the current to AFM layer gives rise
to variation of macroscopic magnetization and, as will be shown
below, induces rotation of the magnetic lattice. R.h.s. of
Eq.~(\ref{eq_dynamics_2}) is thus can be considered as the
Slonczewski's term for STT. In the case of thin AFM layer it can
be expressed in a form Eq.~(\ref{eq_flux_1}) with an obvious
substitution of $\mathbf{M}_{\mathrm{FM}}\rightarrow
\mathbf{M}_{\mathrm{AFM}}$, $S_{\mathrm{FM}}\rightarrow
S_{\mathrm{AFM}}$, $v_{\mathrm{FM}}\rightarrow v_{\mathrm{AFM}}$.
Tensor $\hat \epsilon$ has the symmetry-imposed structure
analogous to that given by Eq.(\ref{eq_Slon_1}). In the simplest
case of spin glasses and some noncollinear AFMs (like FeMn) this
tensor is isotropic, some other particular cases will be discussed
below.

It should be noted that the spin-polarized current affects an AFM
layer in three ways. First of all, due to the flip of the free
electron spins at the NM/AFM interface (or within an AFM volume,
if the thickness of layer is rather small) the current transfers
spin moment and, correspondingly, magnetization to the localized
magnetic lattice (see Eq.(\ref{eq_flux_1})). This effect is
analogous to the Slonczewski's nonadiabatic STT.

 Second, the rest of the current that hasn't change polarization, contributes into
the effective magnetic field due to exchange coupling (effective
constant $I_{\mathrm{sd}}$) between the localized and free
electrons:
\begin{equation}\label{eq_current-field}
\mathbf{H}^{\mathrm{curr}}_{\mathrm{eff}}\propto\hbar\gamma
I_{\mathrm{{sd}}}j(1-2\epsilon_{\mathrm{s-f}})\mathbf{p}_{\mathrm{curr}}
\end{equation}
(similar to adiabatic spin torque that acts on the FM domain
walls, see, e.g. Ref.~\onlinecite{Zhang:PhysRevLett.93.127204}).

At last, the current itself (no matter polarized or no) produces
an Oersted field. The last contribution can be easily taken into account
in the dynamic equations as a component of the external magnetic
field.

The first contribution depends mainly from the scattering
properties of the NM/AFM interface and can be equally strong (or
equally weak) in FMs and AFMs, independently from the magnetic
structure of the soft layer (at least, in the first
approximation). The second contribution can play an important role
in the case of an ac current and possibly triggers the spin
oscillations, in analogy with the effective magnetic field
generated by an inverse Faraday effect.
\cite{Ivanov:2009,Ivanov:PhysRevLett.105.077402}

\subsection{Dynamic equations}
In what follows we consider the case of AFMs and other
multisublattice magnets with strong inter- and intra-sublattice
exchange couplings -- so strong, that the applied magnetic field
and/or high density current do not influence the mutual
orientation of sublattice magnetizations. The dynamics of such a
system can be effectively described as a 3D solid-like rotation of
sublattice magnetizations stocked together by the exchange
interactions or, equivalently, of one,
 two, or three mutually orthogonal vectors that characterize the magnetic structure
 (see Fig.~\ref{fig_geometry} b).
Appropriate and adequate technique for description of dynamics of the magnetic
structure is the Lagrange formalism which makes it possible to
exclude from consideration those degrees of freedoms that are
related with the mutual tilt of sublattice magnetizations (so
called exchange modes). Convenient parametrization and explicit
form of the Lagrange function depends upon the type of magnetic
ordering. In noncollinear magnets the 3D rotation can be
parametrized with the help of the Gibbs' vector
$\boldsymbol\varphi=\varphi \mathbf{e}$, where the unit vector
$\mathbf{e}$ defines an instantaneous rotation axis,
$\varphi=\tan(\theta/2)$, and $\theta$ is the rotation angle (see
Fig.\ref{fig_geometry} b). Generalized coordinates are generated
 by the infinitesimal spin rotation $\delta\boldsymbol\theta$ (see
Appendix \ref{appen_A}) that represents the difference between
rotations
$\mathbf{\boldsymbol\varphi}+d\mathbf{\boldsymbol\varphi}$ and
$\mathbf{\boldsymbol\varphi}$:
\begin{equation}\label{eq_rotation_angle_1}
   \delta\boldsymbol\theta=2\frac{d\boldsymbol\varphi+\boldsymbol\varphi\times d\boldsymbol\varphi}{1+{\boldsymbol\varphi^2}}
   \end{equation}
   or
   \begin{equation}\label{eq_rotation_angle_2}
   \delta\theta_\alpha=2\lambda_{\alpha\beta}d\varphi_\beta,
   \quad\lambda_{\alpha\beta}=\frac{\delta_{\alpha\beta}+\varepsilon_{\alpha\gamma\beta}\varphi_\gamma}{1+{\boldsymbol\varphi^2}},
\end{equation}
where $\varepsilon_{\alpha\gamma\beta}$ is the (completely
antisymmetric)
 Levi-Civita symbol.
The components of spin rotation frequency
$\mathbf{\Omega}\equiv\dot{\boldsymbol\theta}$ form a set of
corresponding generalized velocities and could be expressed
through $\boldsymbol\varphi$ as follows:
\begin{equation}\label{eq_rotation_frequency}
    \mathbf{\Omega}=2\frac{\dot{\boldsymbol\varphi}+\boldsymbol\varphi\times\dot{\boldsymbol\varphi}}{1+{\boldsymbol\varphi^2}}.
\end{equation}

In the presence of the external magnetic field $\mathbf{H}$ the
Lagrange function for AFMs takes the form \cite{Andreev:1980}
\begin{equation}\label{eq_Lagrange_1}
    \mathcal{L}_{\mathrm{AFM}}=\frac{1}{2\gamma^2}\chi_{\alpha\beta}(\boldsymbol\varphi)\left(\Omega_\alpha+\gamma H_\alpha\right)\left(\Omega_\beta+\gamma
    H_\beta\right)-U_{\mathrm{AFM}}(\boldsymbol\varphi),
\end{equation}
where $\chi(\boldsymbol\varphi)$ is the magnetic susceptibility
tensor that accounts for the exchange symmetry of particular AFM.
For the sake of simplicity we consider further the case of the
isotropic media (typical for spin glasses and some noncollinear
AFMs) with
$\chi_{\alpha\beta}(\boldsymbol\varphi)\equiv\chi\delta_{\alpha\beta}$,
$\chi=const$;  generalization on more complicated cases is
straightforward. The symbol $U_{\mathrm{AFM}}(\boldsymbol\varphi)$
is the potential energy that depends upon the magnetic anisotropy
of the soft layer (see Appendix \ref{appen_A}).

In the absence of dissipation (and, in particular, spin-polarized
current) the dynamic equations are deduced from
(\ref{eq_Lagrange_1}) as Euler-Lagrange equations
\begin{equation}\label{eq_Euler-Lagrange}
  \frac{d}{dt}\left(\frac{\partial \mathcal{L}_{\mathrm{AFM}}}{\partial \dot{\boldsymbol
  \varphi}}\right)-\frac{\partial \mathcal{L}_{\mathrm{AFM}}}{\partial \boldsymbol
  \varphi}=0,
\end{equation}
and take a form:
\begin{widetext}
\begin{equation}\label{eq_dynamic_1}
    \frac{2\chi}{\gamma^2}\lambda_{\beta\alpha}\frac{d}{dt}\left[\left(\Omega_\beta+\gamma
    H_\beta\right)\right]+\frac{2}{1+\boldsymbol{\varphi}^2}\left\{(\mathbf{H}\times\dot{\boldsymbol\varphi})_\alpha+H_\beta[\lambda_{\beta\gamma}\dot{\varphi}_\gamma\varphi_\alpha-\lambda_{\beta\alpha}(\boldsymbol{\varphi}\dot{\boldsymbol{\varphi}})]\right\}+\frac{\partial
    U_{\mathrm{AFM}}}{\partial
    \varphi_\alpha} =0.
\end{equation}
\end{widetext} On the other hand, according to
Ref.~\onlinecite{Andreev:1980}, macroscopic magnetization is
proportional to spin rotation frequency:
\begin{equation}\label{eq_magnetization_1}
   \mathbf{M}_{\mathrm{AFM}}=\frac{\chi}{\gamma}\left(\boldsymbol\Omega+\gamma
   \mathbf{H}\right).
\end{equation}
So, multiplying Eq.~(\ref{eq_dynamic_1}) by $\gamma
\lambda^{-1}_{\alpha\delta}$  one can reduce it to a form
\begin{equation}\label{eq_dynamic_4}
\dot{\mathbf{M}}_{\mathrm{AFM}}+\chi\mathbf{H}\times\boldsymbol{\Omega}+\frac{\gamma}{2}\hat{\lambda}^{-1}\cdot\frac{\partial
U_{\mathrm{AFM}} }{\partial\boldsymbol{\varphi}}=0,
\end{equation}
or, equivalently, with account of Eq.~(\ref{eq_magnetization_1}):
\begin{equation}\label{eq_dynamic_4b}
\frac{\partial\mathbf{M}_{\mathrm{AFM}}}{\partial
t}+\underbrace{\gamma\mathbf{H}\times\mathbf{M}_{\mathrm{AFM}}+\frac{\gamma}{2}\hat{\lambda}^{-1}\cdot\frac{\partial
U_{\mathrm{AFM}}}{\partial\boldsymbol{\varphi}}}_{\mathbf{T}_\mathrm{AFM}}=0.
\end{equation}
Here
$\lambda^{-1}_{\alpha\beta}=\delta_{\alpha\beta}+\varepsilon_{\alpha\beta\gamma}\varphi_\gamma+\varphi_\alpha\varphi_\beta$
is the tensor inverse to $\lambda_{\alpha\beta}$. Partial time
derivative is used to stress that the free variables, in addition
to time, include the generalized coordinated
$\boldsymbol{\varphi}$.

It can be easily seen that Eq.~(\ref{eq_dynamic_4b}) looks similar
to Eq.~(\ref{eq_LLG}). The torque $\mathbf{T}_\mathrm{AFM}$ is
produced by the external field (like in FMs) and by the magnetic
anisotropy (similar but not equal to that in FMs). Thus, equations
(\ref{eq_dynamic_4}), (\ref{eq_dynamic_4b}) and hence
(\ref{eq_dynamic_1}) could be treated as the balance equation for
magnetization which in the absence of current has a
form\footnote{~Full time derivative includes
$\dot{\boldsymbol\varphi}(\partial/\partial \boldsymbol\varphi)$.}
\begin{equation}\label{eq_dynamics_3}
   \frac{d\mathbf{M}_{\mathrm{AFM}}}{dt}= 0.
\end{equation}

In the presence of spin-polarized current the balance equation
(\ref{eq_dynamics_3}) transforms into Eq.~(\ref{eq_dynamics_2}).
Then, multiplying the latter by $2\hat\lambda/\gamma$ and taking
into account Eq.~(\ref{eq_flux_1}) we arrive to
\begin{widetext}
\begin{equation}\label{eq_dynamic_5}
   \frac{2\chi\lambda_{\beta\alpha}}{\gamma^2}\frac{d}{dt}\left[\left(\Omega_\beta+\gamma
    H_\beta\right)\right]+\frac{2}{1+\boldsymbol{\varphi}^2}\left\{(\mathbf{H}\times\dot{\boldsymbol\varphi})_\alpha+H_\beta[\lambda_{\beta\gamma}\dot{\varphi}_\gamma\varphi_\alpha-\lambda_{\beta\alpha}(\boldsymbol{\varphi}\dot{\boldsymbol{\varphi}})]\right\}+\frac{\partial
    U_{\mathrm{AFM}}}{\partial
    \varphi_\alpha}=\frac{2}{\gamma}\lambda_{\beta\alpha}gjp^\beta_{\mathrm{curr}},
\end{equation}
\end{widetext}
where we have introduced the phenomenological constant
$g\equiv\hbar\gamma
S_\mathrm{AFM}\epsilon_\mathrm{s-f}/(|e|v_\mathrm{AFM})$.

Eq.~(\ref{eq_dynamic_5}) can be considered as the Lagrange
equation in the presence of the external dissipative forces
\begin{equation}\label{eq_Lagrange}
  \frac{d}{dt}\left(\frac{\partial \mathcal{L}_{\mathrm{AFM}}}{\partial \dot{\boldsymbol
  \varphi}}\right)-\frac{\partial \mathcal{L}_{\mathrm{AFM}}}{\partial \boldsymbol
  \varphi}=-\frac{\partial\mathcal{R}_{\mathrm{ AFM}}}{\partial \dot{\boldsymbol
  \varphi}},
\end{equation}
where $\mathcal{R}_{\mathrm{ AFM}}$ is the Rayleigh dissipation
function, $\mathcal{R}_{\mathrm{ AFM}}$ is related with the rate
of the energy $\mathcal{E}_{\mathrm{AFM}}$ losses as follows:
\begin{equation}\label{eq-energy_losses}
  \frac{d\mathcal{E}_{\mathrm{AFM}}}{dt}\equiv-\dot{\boldsymbol
  \varphi}\left(\frac{\partial \mathcal{R}_{\mathrm{ AFM}}}{\partial\dot{\boldsymbol
  \varphi}}\right).
\end{equation}
Comparison of Eqs.~(\ref{eq_Lagrange}) and (\ref{eq_dynamic_5})
with account of expressions (\ref{eq_rotation_angle_2}) and
(\ref{eq_rotation_frequency}) shows that the current-induced
contribution into dissipation function can be presented as
$-gj(\mathbf{\Omega}\cdot\mathbf{p}_{\mathrm{curr}})/\gamma$.

The internal losses (Gilbert damping) in AFM layer is taken into
account in a standard way. As a result, dissipation function takes
a form
\begin{equation}\label{eq_Rayleigh_2}
\mathcal{R}_{\rm
AFM}=\frac{\alpha_{\mathrm{AFM}}}{2\gamma}\mathbf{\Omega}^2-\frac{gj}{\gamma}
  (\mathbf{\Omega}\cdot\mathbf{p}_{\mathrm{curr}}),
\end{equation}
where $\alpha_{\mathrm{AFM}}$ is a damping constant of AFM layer.

Expression (\ref{eq_Rayleigh_2}) can be written in terms of the
Gibbs vector as follows (see Eq.~(\ref{eq_rotation_frequency})):
\begin{widetext}
\begin{equation}\label{eq_Rayleigh_3}
\mathcal{R}_{\rm
AFM}=\frac{2\alpha_{\mathrm{AFM}}}{\gamma}\left[\frac{\dot{\boldsymbol\varphi}^2}{1+{\boldsymbol\varphi^2}}-\frac{(\boldsymbol\varphi\cdot\dot{\boldsymbol\varphi})^2}{(1+{\boldsymbol\varphi^2})^2}\right]-\frac{2gj}{\gamma}
  \frac{\mathbf{p}_{\mathrm{curr}}\cdot\dot{\boldsymbol\varphi}+\mathbf{p}_{\mathrm{curr}}\cdot(\boldsymbol\varphi\times\dot{\boldsymbol\varphi})}{1+{\boldsymbol\varphi^2}},
\end{equation}
\end{widetext}

The expression (\ref{eq_Rayleigh_3}) for the Rayleigh function is
the main result of the present paper. Together with the Lagrange
function (\ref{eq_Lagrange_1}) it describes the dynamics of AFMs
with strong exchange coupling in the presence of spin-polarized
current.

As it was already shown, expressions (\ref{eq_Rayleigh_2}),
(\ref{eq_Rayleigh_3}) for the Rayleigh function result from the
spin conservation principle and thus are rather general. However,
the analogous expressions could be deduced directly from the
Landau-Lifshits-Slonczewski equations for the magnetic sublattices
with the account of strong exchange coupling between them.
\cite{gomo:2010} Moreover, for the case of FMs the current-induced
contribution into the Rayleigh function has a similar form (see
Ref. \onlinecite{Ivanov:PhysRevLett.99.247208}):
\begin{equation}\label{eq_Rayleigh_FM}
    \mathcal{R}_{\rm FM}=\ldots-\frac{gj}{\gamma M_0^2}
  \mathbf{p}_{\mathrm{curr}}\cdot(\mathbf{M}\times\dot\mathbf{M})
\end{equation}
where by $\ldots$ we denote the contribution from the internal
damping.

Analysis of the expression (\ref{eq_Rayleigh_2})
 shows that the stationary state of AFMs
(with $d\mathcal{E}_{\mathrm{AFM}}/dt=0$) in the presence of
steady spin-polarized current corresponds to rotation of the
magnetic structure around the current polarization
$\mathbf{p}_{\mathrm{curr}}$ with the constant frequency
\begin{equation}\label{eq_frequency}
\mathbf{\Omega}=\frac{g}{2\alpha_{\mathrm{AFM}}}j\mathbf{p}_{\mathrm{curr}}. 
\end{equation}
The rotation frequency can be controlled by the current value $j$
and depends upon the loss factor $\alpha_{\mathrm{AFM}}$.

\section{Dissipation function for different types of
antiferromagnets}\label{sec_different_AFM} The method described in
the Sec.\ref{sec_Lagrange} can be easily generalized to the
systems of different exchange symmetry. In this section we analyze
some typical cases useful for applications. For the sake of
clarity we reproduce the expressions for the kinetic energy from
Ref.~\onlinecite{Andreev:1980} as well.

\subsection{Noncollinear antiferromagnets and disordered magnets}
\begin{figure}[htbp]
 \includegraphics[width=1\columnwidth]{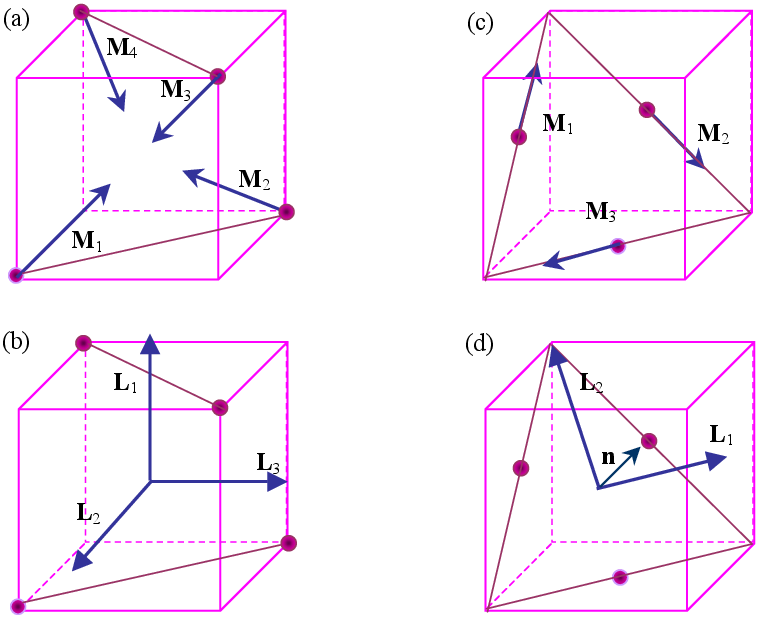}
  \caption{(Color online) \textbf{Magnetic structure of the cubic noncollinear AFMs}.
  (a) Magnetic structure of the disordered FeMn. Magnetic atoms
  (circles) form fcc lattice, vectors $\mathbf{M}_k$  of sublattice magnetizations
  (arrows) point to $\langle 111\rangle$ directions. (b) This
  structure can be equivalently described in terms of three mutually orthogonal AFM
  vectors $\mathbf{L}_k$ (see Eq.~(\ref{eq_FeMn_AFM_vectors})). (c)  Planar magnetic structure of IrMn$_{3}$
  and some of the antiperovskites Mn$_3$MN, where M=Ni, Ga, Ag, Zn. Magnetic moments $\mathbf{M}_k$
  (arrows) are localized on Mn atoms (circles). (d) Planar
  magnetic structure can be described by two mutually orthogonal AFM
  vectors $\mathbf{L}_k$ (see Eq.~(\ref{eq_IrMn_AFM_vectors})). The plane of AFM ordering is
  described by the unit vector  $\mathbf{n}\|\mathbf{L}_1\times\mathbf{L}_2$. \label{fig_magnetic_lattice}}
\end{figure}

We start from the simplest case of noncollinear AFMs with the
cubic exchange symmetry. A typical example is given by the
metallic AFM FeMn widely used in spin-valves. Magnetic structure
of FeMn (see Fig.~\ref{fig_magnetic_lattice} a,b) consists of four
equivalent magnetic sublattices $\mathbf{M}_k$ ($k=1,\ldots, 4$)
oriented along four $\langle 111\rangle$ directions of cubic cell.
\cite{Endoh:1971} 
Macroscopic magnetization
($\mathbf{M}_{\mathrm{AFM}}=\sum\mathbf{M}_k$) of FeMn in
equilibrium is zero, so, the magnetic structure is described by
three orthogonal AFM vectors that could be related with the
sublattice magnetizations as follows:
\begin{eqnarray}\label{eq_FeMn_AFM_vectors}
\mathbf{L}_1&=&\mathbf{M}_1+\mathbf{M}_2-\mathbf{M}_3-\mathbf{M}_4,\nonumber\\
\mathbf{L}_2&=&-\mathbf{M}_1+\mathbf{M}_2-\mathbf{M}_3+\mathbf{M}_4,\nonumber\\
\mathbf{L}_3&=&\mathbf{M}_1-\mathbf{M}_2-\mathbf{M}_3+\mathbf{M}_4.
\end{eqnarray}
It is interesting that such a complicated structure is predicted not
only in the bulk material, but also in the thin (down to 1.5 nm \cite{Jungblut:1994, Kuch:2002(2)}) FeMn layer  within multilayered structures used 
in spintronic devices. \cite{Urazhdin:2007, Canning:2002}

All three  vectors $\mathbf{L}_k$ have the same modulus, so,
magnetic structure in the exchange approximation has a cubic
symmetry and tensors $\hat\epsilon$ and $\hat\chi$ are isotropic.
Thus, the Rayleigh function is given by the
Exp.~(\ref{eq_Rayleigh_3}) and the kinetic energy has a form:
\begin{equation}\label{eq_kinetic_1}
T_{\mathrm{kin}}=\frac{2\chi}{\gamma^2}\left[\frac{\dot{\boldsymbol\varphi}^2}{1+{\boldsymbol\varphi^2}}-\frac{(\boldsymbol\varphi\cdot\dot{\boldsymbol\varphi})^2}{(1+{\boldsymbol\varphi^2})^2}\right]+\frac{2\chi}{\gamma}\frac{\mathbf{H}\cdot\dot{\boldsymbol\varphi}+\mathbf{H}\cdot(\boldsymbol\varphi\times\dot{\boldsymbol\varphi})}{1+{\boldsymbol\varphi^2}}.
\end{equation}
Similar expressions for the kinetic energy and Rayleigh function
are obtained for the disordered systems that according to
Ref.~\onlinecite{Andreev:1980} include spin glasses (no
macroscopic magnetization)  and disordered
AFMs\footnote{~Disordered magnetics also include FM materials with
the nonzero magnetization, however, here we are interested only in
the materials with zero or small macroscopic magnetic moment.}
(zero macroscopic magnetization and nonzero AFM vectors). These
systems show strong magnetic correlations induced by exchange
coupling between the nearest localized spins and are isotropic
with respect to any spin
rotations.
\subsection{Planar noncollinear antiferromagnets}
Noncollinear AMFs with triangular magnetic structure (see
Fig.~\ref{fig_magnetic_lattice}c,d) are interesting and important
for applications. In particular, such type of structure is
observed in IrMn$_3$ alloy \cite{szunyogh-2009-79} which is, as
well as FeMn, used as a pinning layer in spintronic devices (see,
e.g., Ref.\onlinecite{Bass:2010,King:0022-3727-34-4-315}), and in
the series of metallic antiperovskites Mn$_3$MN (where M=Ni, Ga,
Ag, Zn) \cite{Fruchart:1971} used in composite materials due to
strong negative thermal expansion (see, e.g.
Refs.~\onlinecite{Takenaka:2005, Takenaka:2006,
Takenaka:2010PhRvB81v4419K}). Some of the antiperovskites also
show giant magnetoresistance effect
\cite{Kamishima:PhysRevB.63.024426} in AFM phase and thus can be
susceptible to the action of spin-polarized current.

Magnetic structure of cubic IrMn$_3$ and Mn$_3$MN is represented
by three equivalent magnetic sublattices with magnetization
vectors $\mathbf{M}_k$, ($k=1,2,3$) that make 120$^\circ$ angle
with respect to each other. Two mutually orthogonal AFM vectors,
\begin{equation}\label{eq_IrMn_AFM_vectors}
\mathbf{L}_1=\mathbf{M}_1+\mathbf{M}_2-2\mathbf{M}_3,\quad
\mathbf{L}_2=\sqrt{3}\left(\mathbf{M}_1-\mathbf{M}_2\right),
\end{equation}
define the plane of the magnetic ordering (in exchange
approximation) with the normal vector
$\mathbf{n}\|\mathbf{L}_1\times\mathbf{L}_2$, and equilibrium
magnetization $\mathbf{M}_{\mathrm{AFM}}=\sum\mathbf{M}_k=0$.

Exchange symmetry group is isomorphous to $C_{3h}$, where rotation
axis is parallel to $\mathbf{n}$, so, any 2-nd rank symmetric
tensor has two independent components -- parallel and
perpendicular to $\mathbf{n}$ direction. For example, tensor of
magnetic susceptibility takes a form:
\begin{equation}\label{eq_magnetic_susceptibility}
 \hat \chi=\chi_\perp\hat
 1+\left(\chi_\|-\chi_\perp\right)\mathbf{n}\otimes\mathbf{n}.
\end{equation}
 As a result, the kinetic energy and Rayleigh function can be
written as:
\begin{eqnarray}\label{eq_kinetic_2}
T_{\mathrm{kin}}&=&\frac{\chi_\perp}{2\gamma^2}(\mathbf{\Omega}+\gamma
\mathbf{H})^2+\frac{\chi_\|-\chi_\perp}{2\gamma^2}\left[(\mathbf{\Omega}+\gamma
\mathbf{H})\cdot\mathbf{n}\right]^2,\nonumber\\
\mathcal{R}_{\mathrm{AFM}}&=&\frac{\alpha_\perp}{2\gamma}\mathbf{\Omega}^2+\frac{\alpha_\|-\alpha_\perp}{2\gamma}\left(\mathbf{\Omega}\cdot\mathbf{n}\right)^2-\frac{g_{\perp}j}{\gamma}(\mathbf{\Omega}\cdot\mathbf{p}_{\mathrm{curr}})\nonumber\\
&-&\frac{\left(g_\|-g_{\perp}\right)j}{\gamma}(\mathbf{\Omega}\cdot\mathbf{n})(\mathbf{n}\cdot\mathbf{p}_{\mathrm{curr}}).
\end{eqnarray}
The functions $T_{\mathrm{kin}}$ and $\mathcal{R}_{\mathrm{AFM}}$
from the above expression could be easily expressed in terms of
the Gibbs' vector using the relation
(\ref{eq_rotation_frequency}).
\subsection{Collinear antiferromagnets}
The collinear AFMs with high Neel temperature, like IrMn and NiO,
are also widely used in spin-valves. Their magnetic structure can
be described with the only AFM vector $\mathbf{L}$ with the fixed
modulus. Convenient parametrization for rotation of this vector
includes only two independent variables. However, times
derivative, $\dot{\mathbf{L}}=\mathbf{\Omega}\times\mathbf{L}$, is
expressed through the rotation vector. This means that for the
tensors $\hat\chi$, $\hat \alpha$  and $\hat g$ the component
parallel to $\mathbf{L}$ is zero, e.g. (compare with
Eq.~(\ref{eq_Slon_1})):
\begin{equation}\label{eq_magnetic_susceptibility_2}
 \hat \chi=\chi_\perp\left(\hat
 1-\mathbf{e}_{\mathbf{L}}\otimes\mathbf{e}_{\mathbf{L}}\right),\quad
 \mathbf{e}_{\mathbf{L}}\equiv\frac{\mathbf{L}}{|\mathbf{L}|}.
\end{equation}

The kinetic energy and Rayleigh function take a form:
\begin{eqnarray}\label{eq_kinetic_3}
T_\mathrm{kin}&=&\frac{\chi_\perp}{2\gamma^2}\left\{(\mathbf{\Omega}+\gamma
\mathbf{H})^2-\left[(\mathbf{\Omega}+\gamma
\mathbf{H})\cdot\mathbf{e}_{\mathbf{L}}\right]^2\right\},\nonumber\\
\mathcal{R}_{\mathrm{AFM}}&=&\frac{\alpha_\perp}{2\gamma}\left[\mathbf{\Omega}^2-\left(\mathbf{\Omega}\cdot\mathbf{e}_{\mathbf{L}}\right)^2\right]-\frac{g_{\perp}j}{\gamma}(\mathbf{\Omega}\cdot\mathbf{p}_{\mathrm{curr}})\nonumber\\
&+&\frac{g_{\perp}j}{\gamma}(\mathbf{\Omega}\cdot\mathbf{e}_{\mathbf{L}})(\mathbf{e}_{\mathbf{L}}\cdot\mathbf{p}_{\mathrm{curr}}),
\end{eqnarray}
in correspondence with the expressions deduced in Refs.
\onlinecite{Bar-june:1979E} and \onlinecite{gomo:2010}.
\section{Current-induced instability}
To illustrate an application of the obtained expressions for the
Rayleigh function, we consider small oscillations of AFM vectors
in the presence of dc and ac spin-polarized current (external
magnetic field $\mathbf{H}_{\mathrm{ext}}=0$). We start from the
simplest case of isotropic noncollinear AFM. Linearized equations
for $\boldsymbol\varphi$ components are derived from the
Exp.~(\ref{eq_kinetic_1}) and (\ref{eq_Rayleigh_3}) as follows:
\begin{eqnarray}\label{eq_noncollinear_frequency}
\ddot{\boldsymbol\varphi}&+&\frac{\gamma
\alpha_{\mathrm{AFM}}}{\chi}\dot{\boldsymbol\varphi}+\gamma\beta
j\mathbf{p}_{\mathrm{curr}}\times\dot{\boldsymbol\varphi}+\frac{\gamma}{2}\left(\beta\frac{dj}{dt}-\frac{gj}{\chi}\right)\mathbf{p}_{\mathrm{curr}}\times\boldsymbol\varphi
\nonumber \\ &+&\omega_\mathrm{AFMR}^2\boldsymbol\varphi
=\frac{\gamma}{2}\left(\frac{gj}{\chi}-
\beta\frac{dj}{dt}\right)\mathbf{p}_{\mathrm{curr}}.
\end{eqnarray}
Here we take into account an ``adiabatic'' torque that acts as an
effective field (\ref{eq_current-field}) and is expressed as
$\mathbf{H}^{\mathrm{curr}}_{\mathrm{eff}}\equiv\beta
j\mathbf{p}_{\mathrm{curr}}$ (where $\beta$ is the phenomenologic
constant that depends upon $sd$-exchange ), $\omega_\mathrm{AFMR}$
is the 3-times degenerated AFMR frequency in the absence of field
and current.

\begin{figure}[htbp]
  \includegraphics[width=1\columnwidth]{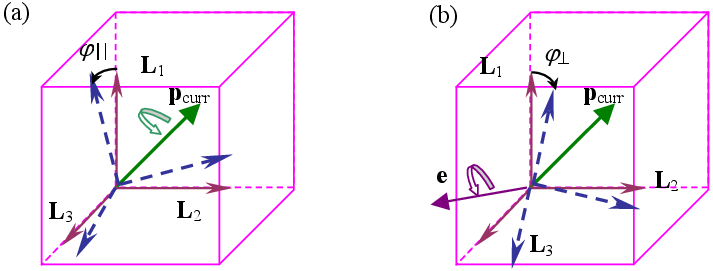}
  \caption{(Color online) \textbf{Current-induced rotation of AFM structure}. (a) Magnetic structure represented by three AFM vectors
  $\mathbf{L}_k$ is rotated through the fixed angle
  $\varphi_\|$ under the action of dc current. Rotation axis
  $\mathbf{e}$ (not shown) is parallel to the vector of current
  polarization $\mathbf{p}_{\mathrm{curr}}$. (b) Overcritical ($J\ge
  J^{\mathrm{AFM}}_{\mathrm{cr}}$) current induces instability
  with respect to rotation around the axis $\mathbf{e}\perp\mathbf{p}_{\mathrm{curr}}$. \label{fig_rotation}}
\end{figure}

Analysis of Eq.~(\ref{eq_noncollinear_frequency}) reveals some
interesting features of spin-torque phenomena in AFMs. First, the
current induces small rotation (dc) or small oscillations (ac) of
the magnetic sublattices around $\mathbf{p}_{\mathrm{curr}}$
direction (see Fig.~\ref{fig_rotation} a). Taken, for example,
$j=j_0e^{i\omega t}$, one obtains from
(\ref{eq_noncollinear_frequency}) for the component
$\varphi_\|=\boldsymbol\varphi\cdot\mathbf{p}_{\mathrm{curr}}$:
\begin{equation}\label{eq_parallel_rotation}
  \varphi_\|=\frac{\gamma
  j_0\left(g-i\omega
\chi\beta\right)}{2\chi\left[\omega_\mathrm{AFM}^2-\omega^2+2i\gamma_{\mathrm{AFM}}\omega\right]},
\end{equation}
where $2\gamma_{\mathrm{AFM}}\equiv\gamma\alpha_\mathrm{AFM}/\chi$
is the half-width of AFM resonance. Current density $j_0$ is
supposed to be rather small so, that the approximation
$\varphi_\|\ll 1$ is valid.

It is easy to see from Eq.~(\ref{eq_parallel_rotation}) that the
ac current transfers both nonadiabatic (term with $g$) and
adiabatic (term with $\beta$) spin torques to AFM layer, while dc
current ($\omega=0$) transfers nonadiabatic torque only. The ac
current induces the resonance at $\omega=\omega_\mathrm{AFM}$. The
dc current induces small deflection of AFM vectors from
equilibrium direction:
$\mathbf{L}_k-\mathbf{L}_k^{(0)}=2\varphi_\|\mathbf{p}_{\mathrm{curr}}\times
\mathbf{L}_k^{(0)}$.

Second, nonadiabatic (Slonczewski's) contribution can induce
reorientation of the magnetic structure seen as instability with
respect to rotation around the axis
$\mathbf{e}\perp\mathbf{p}_{\mathrm{curr}}$ (see
Fig.~\ref{fig_rotation} b). The value of critical current at which
such an instability takes place, $J^{\mathrm{AFM}}_{\mathrm{cr}}$,
depends on the damping coefficient, in analogy with FM:
\begin{equation}\label{eq_critical current_AFM}
J^{\mathrm{AFM}}_{\mathrm{cr}}=\chi\frac{4\gamma_{\mathrm{AFM}}\omega_{\mathrm{AFMR}}}{g\gamma}
S_{\mathrm{AFM}}=\chi \frac{2\omega^2_{\mathrm{AFMR}}}{g\gamma
Q_{\mathrm{AFM}}}S_{\mathrm{AFM}},
\end{equation}
where $Q_{\mathrm{AFM}}\equiv
\omega_{\mathrm{AFMR}}/(2\gamma_{\mathrm{AFM}})$ is a quality
factor of AFMR. However, there is substantial difference between
the critical current in AFM and corresponding value for FM
\cite{Ivanov:PhysRevLett.99.247208}:
\begin{equation}\label{eq_critical_current_FM}
    J^{\mathrm{FM}}_{\mathrm{cr}}\propto \frac{2\omega^2_{\mathrm{FMR}}}{g\gamma
Q_{\mathrm{FM}}}S_{\mathrm{FM}}.
\end{equation}

Really, in AFMs the critical value
$J^{\mathrm{AFM}}_{\mathrm{cr}}$ is proportional to the magnetic
susceptibility $\chi$ which in AFMs is rather small (compared to
FM materials). So, for FMs and AFMs with equal quality factors and
comparable efficiency $\epsilon_{\mathrm{s-f}}$ of spin-flip
scattering the critical value of current for AFMs
$J^{\mathrm{AFM}}_{\mathrm{cr}}\propto
J^{\mathrm{FM}}_{\mathrm{cr}}\chi\ll
J^{\mathrm{FM}}_{\mathrm{cr}}$ can be significantly smaller than
in FMs (due to the so called effect of the exchange enhancement).

Similar effects are predicted for the planar and collinear AFMs.
The planar AFM has two different AFMR frequencies
$\omega_{\mathrm{AFMR}}^\|<\omega_{\mathrm{AFMR}}^\perp$
corresponding to the in-plane and out-of-plane rotations of the
magnetic vectors ($\omega_{\mathrm{AFMR}}^\perp$ is double
degenerated). In this case the current-induced dynamics depends
upon the orientation of the current polarization vector
$\mathbf{p}_{\mathrm{curr}}$ with respect to the plane formed by
AFM vectors (defined by the normal vector $\mathbf{n}$). If
$\mathbf{p}_{\mathrm{curr}}\|\mathbf{n}$, then the current induces
small rotations or oscillations around
$\mathbf{p}_{\mathrm{curr}}$ with the resonant frequency
$\omega_{\mathrm{AFMR}}^\|$ (instead of $\omega_{\mathrm{AFMR}}$
in (\ref{eq_parallel_rotation})), in analogy with the previous
case of isotropic AFMs.

If the vector $\mathbf{p}_{\mathrm{curr}}\perp\mathbf{n}$, then
the value of the critical current at which the  AFM vectors (and,
correspondingly, $\mathbf{n}$) starts to rotate around
$\mathbf{e}\|\mathbf{p}_{\mathrm{curr}}\times\mathbf{n}$ (see
Fig.~\ref{fig_rotation}) depends upon the magnetic anisotropy of
the system, namely,
\begin{widetext}
\begin{equation}\label{eq_critical current_AFM_2}
J^{\mathrm{AFM}}_{\mathrm{cr}}=\frac{2\chi S_{\mathrm{AFM}}
}{g\gamma}\sqrt{2\gamma^2_{\mathrm{AFM}}\left[\left(\omega_{\mathrm{AFMR}}^\|\right)^2+\left(\omega_{\mathrm{AFMR}}^\perp\right)^2\right]+\left[\frac{\left(\omega_{\mathrm{AFMR}}^\|\right)^2-\left(\omega_{\mathrm{AFMR}}^\perp\right)^2}{4}\right]^2}.
\end{equation}
\end{widetext}
This result is similar to the critical current obtained for the
collinear AFMs.\cite{gomo:2010}

It should be stressed that in contrast to FMs, the critical
current is still finite even for zero dissipative constant and is
dictated by magnetic anisotropy (the second term under the square
root in Eq.~(\ref{eq_critical current_AFM_2})). This nontrivial
effect is peculiar to AFMs and can be explained as follows.
Dynamic magnetization $\mathbf{M}_{\mathrm{AFM}}$, as seen from
Eq.~(\ref{eq_magnetization_1}), is related with the rotation of
AFM vectors and hence, with circularly polarized modes of magnetic
excitations  (we consider only the long wave modes). Moreover, for
linearly polarized oscillations of AFM vectors
$\mathbf{M}_{\mathrm{AFM}}=0$. So, the spin-polarized current
pumps energy into circularly polarized modes and does not interact
with the linear ones. On the other hand, polarization of eigen
modes depends upon the magnetic anisotropy and in the case under
consideration is linear. Spin-polarized current affects not only
the dissipation but the values of eigen frequencies as well. At
the critical current value (\ref{eq_critical current_AFM_2})) the
splitting between the eigen frequencies is removed and circularly
polarized modes start to ``take energy'' from the current.

Thus, current-induced dynamics of AFMs has some similarities with
the dynamics of FMs, but is richer even in linear approximation.
Destabilization of AFM structure (with respect to solid-like
rotations) takes place at the current values that could be small
compared to FMs.

However, for AFMs with four or more sublattices, the dynamic
equations (\ref{eq_noncollinear_frequency}) describe only three
low frequency modes. Other, so-called exchange (usually high
frequency) modes are related with the mutual tilt of magnetic
moments and could not be described within the present approach.

\section{Detection of spin transfer torque and spin-diode effect}
Up to date the possibility to measure directly STT effects in AFMs
is under discussion. Standard technique for detection of STT
phenomena in the systems with active FM layer is measuring of
magnetoresistance which depends upon mutual orientation of
magnetizations in soft and hard layers. In this section we argue
that the same approach can, in principle, be applied to the
systems with the soft AFM layer.

It follows from (\ref{eq_magnetization_1}) and
(\ref{eq_parallel_rotation}) that current-induced rotation of AFM
vectors produces oscillating macroscopic magnetization:
\begin{equation}\label{eq_parallel_Magnetization }
  \mathbf{M}_{\mathrm{AFM}}=\frac{\omega j_0
  \sqrt{g^2+\chi^2\beta^2\omega^2}
}{2\sqrt{\left(\omega_{\mathrm{AFMR}}^2-\omega^2\right)^2+4\gamma^2_{\mathrm{AFM}}\omega^2}}\cos(\omega
t+\phi)\mathbf{p}_{\mathrm{curr}}.
\end{equation}
where
\begin{equation}\label{eq_phase_shift}
  \phi\equiv\arctan\frac{g(\omega_{\mathrm{AFMR}}^2-\omega^2)-\omega^2\chi\beta\gamma_{\mathrm{AFM}}}{\omega\left[2g\gamma_{\mathrm{AFM}}+\chi\beta(\omega_0^2-\omega^2)\right]}
\end{equation}
is the phase shift between the current and magnetization.

On the other hand, anisotropic magnetoresistance (AMR) of the
multilayer, $\Delta R_{\mathrm{AMR}}$ depends upon the mutual
orientation of oscillating vector, $\mathbf{M}_{\mathrm{AFM}}(t)$,
and fixed magnetization of FM layer,
$\mathbf{M}_\mathrm{pol}\|\mathbf{p}_{\mathrm{curr}}$:
\begin{equation}\label{eq_AMR}
\Delta
R_{\mathrm{AMR}}\propto\mathbf{M}_{\mathrm{AFM}}\cdot\mathbf{p}_{\mathrm{curr}}\propto
j_0\cos(\omega t+\phi).
\end{equation}
This means that:
\begin{itemize}
  \item AMR itself can be used for detection of
spin-torque effect in AFM layer. In the case of multilayers with
the magnetic tunnel junction (instead of metallic nonmagnetic
layer) $\Delta R_{\mathrm{AMR}}$ can be as large as 130 \% (see,
e.g. Ref.~\onlinecite{Park:2010arXiv1011.3188P});
  \item  AMR oscillating with the same frequency as ac current can cause frequency
mixing and a directly measurable dc voltage:
\begin{equation}\label{eq_dc_voltage}
  V_{\mathrm{dc}}=\langle\Delta
R_{\mathrm{AMR}}J\rangle\propto \cos\phi.
\end{equation}
An analogous spin-diode effect in spin-valves with FM soft layer
was already detected in Refs.\onlinecite{Tulapurkar:2005,
Fang:2010arXiv1012.2397F};
  \item The abovementioned frequency mixing can reveal itself in
  the second-harmonic generation with corresponding voltage given by:
  \begin{eqnarray}\label{eq_sh_voltage}
  V_{\mathrm{s.h.}}&=&\Delta
R_{\mathrm{AMR}}J\\ &\propto &\frac{\omega j^2_0
  \sqrt{g^2+\chi^2\beta^2\omega^2}
}{4\sqrt{\left(\omega_{\mathrm{AFM}}^2-\omega^2\right)^2+4\gamma^2_{\mathrm{AFM}}\omega^2}}\cos(2\omega
t+\phi).\nonumber
\end{eqnarray}
\end{itemize}
 In addition, frequency dependence of
$V(\omega)$ can be used for measuring of the ratio between the
dissipative (constant $g$) and nondissipative (constant $\beta$)
torques in AFM. Really, in the vicinity of AFMR
($|\omega-\omega_{\mathrm{AFMR}}|\ll\omega_{\mathrm{AFMR}}$), as
follows from Eqs.~(\ref{eq_AMR}), (\ref{eq_dc_voltage}), and
(\ref{eq_sh_voltage}),
\begin{eqnarray}\label{eq_near_resonance}
 V_{\mathrm{dc}}&\propto&\frac{1}{\sqrt{1+\omega^2(\chi\beta/g)^2}},\nonumber\\
  V_{\mathrm{s.h.}}&\propto &\sqrt{1+\omega^2(\chi\beta/g)^2}.
\end{eqnarray}
Thus, AMR and voltage are observables that provide a direct probe
of the amplitude and phase of the precession of AFM vectors with
respect to the ac current and make it possible to detect
current-induced phenomena in AFM layer.

Expression (\ref{eq_dc_voltage}) shows that spin-diode effect can
be observed in the systems where an AFM plays a role of the soft
magnetic layer. In contrast to FM, where spin-diode effect is
observed in GHz range, \cite{Tulapurkar:2005} AFM layer can
rectify the current in a higher frequency range  (up to THz which
corresponds to AFM resonance).
\section{Discussion and conclusions}
In the present paper we have developed a general phenomenological
theory of current-induced dynamics for AFM with strong exchange
coupling and have derived an expression (\ref{eq_Rayleigh_2}) for
the Rayleigh function  in the presence of spin-polarized current.
We have shown that spin-polarized current can produce a work over
an AFM despite of the absence of equilibrium macroscopic
magnetization. So, spin-polarized current can induce rotation of
the magnetic moments not only in FM, but in any material which
shows magnetic ordering of the exchange nature. The characteristic
values of critical current depend on the efficiency of spin-flip
processes at AFM/NM/FM interfaces but do not depend on the value
of macroscopic magnetization of the soft layer.

Expression (\ref{eq_Rayleigh_2}), in fact, defines the power of
the external current-induced force that acts on the magnetic
system at a given value of the magnetic flux $\boldsymbol
\Phi=\nabla\cdot\hat{\boldsymbol\Pi} \propto
j\mathbf{p}_{\mathrm{curr}}$ flowing into AFM layer (see
Eq.~(\ref{eq_flux_1})). It can be extended to the case of
inhomogeneous distribution of AFM vectors (e.g., in the domain
wall) in the rpesence of steady current as follows:
\begin{eqnarray}\label{eq_Rayleigh_inhomogeneous}
\mathcal{R}_{\mathrm{AFM}}&=&\ldots- \int
\nabla\cdot\hat{\boldsymbol\Pi}\cdot\boldsymbol\Omega
d\mathbf{r}\nonumber\\ &=&\ldots+\frac{d}{dt}\int
\hat{\boldsymbol\Pi}:\left(\nabla\otimes\boldsymbol\theta\right)
d\mathbf{r}\\ & = &\ldots+\frac{d}{dt}\int
\Pi_{lk}\nabla_l\theta_k d\mathbf{r}\nonumber\\
&=&\ldots+\frac{d}{dt}\int \Pi_{lk}\delta\theta_k n_ldS.\nonumber
\end{eqnarray}
Here we denoted with $\ldots$ the terms that do not depend on
current.

 The last equality in Eq.~(\ref{eq_Rayleigh_inhomogeneous})
shows that the magnetic flux density flowing in $\mathbf{n}$
direction, $n_l\Pi_{lk}$, and the rotational vector
$\boldsymbol\theta$ that defines the deflection from the initial
(current-free) state are thermodynamically conjugated variables.
Corresponding elementary work produced by the spin polarized
current at fixed $\nabla\cdot\hat{\boldsymbol\Pi}$ is $\delta
A=\mathbf{n}\cdot\hat{\boldsymbol\Pi}\delta\boldsymbol\theta$.
This means that in general, a steady current-transferred magnetic
flux can change equilibrium orientation of AFM vectors (see, e.g.,
Eq.~(\ref{eq_parallel_rotation}) for the case of dc current).
Inversely, variation of AFM vectors (e.g. in the domain wall)
should give rise to variation of the magnetic flux
$\delta\hat{\boldsymbol\Pi}$. This, in turn, means that the
spin-flux transferred by the current through the AFM domain wall
or other area with inhomogeneous distribution of AFM vectors will
be changed due to, e.g. depolarization. This effect can be, in
principle, detected experimentally, e.g., by measuring
magnetoresistance of FM/AFM/FM multilayer.

Finally, the above approach combined with the principles of
nonequilibrium thermodynamics makes it possible to derive a
general expression for the current-induced non-adiabatic STT for
the bulk conducting AFMs. In order to demonstrate this fact we
consider inhomogeneous distribution of AFM vectors described by
the space dependent rotations $\boldsymbol{\varphi}(\mathbf{r})$
with respect to some reference state. Then, the vector
$\boldsymbol\Omega$ is a generalized flux which is generated by
thermodynamic variable $\boldsymbol{\varphi}$. Corresponding
generalized forces could be obtained from the dynamic equation
(see Eq.~(\ref{eq_dynamic_1}) added with the magnetic damping) for
the electrically open AFM layer in the overdamped regime
($\dot{\boldsymbol\Omega}\rightarrow 0$):
\begin{equation}\label{eq_overdamped_regime}
  \alpha_{\mathrm{AFM}}\boldsymbol{\Omega}-\mathbf{T}_{\mathrm{AFM}}=0,  
\end{equation}
where the torque $\mathbf{T}_{\mathrm{AFM}}$ is defined by the
magnetic anisotropy of AFM layer (see Eq.~(\ref{eq_dynamic_4b})
and explanation below).

 The charge current density $\mathbf{j}$ injected into AFM and the
electric field $\mathbf{E}$ are the other pair of the conjugated
generalized thermodynamic variables. According to the Onsager
principles, generalized fluxes $\boldsymbol\Omega$, $\mathbf{j}$
and generalized forces $\mathbf{T}_{\mathrm{AFM}}$, $\mathbf{E}$
are related as follows:
\begin{equation}\label{eq-Onsager_relations_1}
\boldsymbol\Omega=\hat{L}_{ll}\mathbf{T}_{\mathrm{AFM}}+\hat{L}_{lq}\mathbf{E},\quad\mathbf{j}=\hat{L}_{ql}\mathbf{T}_{\mathrm{AFM}}+\hat{L}_{qq}\mathbf{E},
\end{equation}
where the phenomenological Onsager's coefficients $\hat{L}_{\delta
\xi}$, $\delta \xi=l,q$ should satisfy the rotational symmetry
requirements and reciprocity relations. In assumption of the
electrically homogeneous medium, $\hat{L}_{qq}=\sigma\hat I$,
where $\sigma$ is conductivity, $\hat I$ is a unit matrix, and
$\hat{L}_{ql}=\eta\nabla\otimes \boldsymbol{\theta}$, where $\eta$
is a material constant. From Eq.~(\ref{eq_overdamped_regime}) it
also follows that
$\hat{L}_{ll}=\hat I/\alpha_{\mathrm{AFM}}$

With account of time-inversion symmetry of AFM the reciprocity
relations state that $\hat{L}_{ql}=-\hat{L}_{lq}$ if all the
equilibrium  magnetic vectors of AFM change sign under the time
reversal.

Thus, after quite simple mathematics we get from
Eq.~(\ref{eq-Onsager_relations_1})
\begin{equation}\label{eq-Onsager_relations_2}
\boldsymbol\Omega=\frac{1}{\alpha_{\mathrm{AFM}}}\mathbf{T}_{\mathrm{AFM}}-\frac{\eta}{\sigma}(\mathbf{j},\nabla)\boldsymbol{\theta}.
\end{equation}
Last term in the expression (\ref{eq-Onsager_relations_2})
generalizes the expression for the dissipative torque obtained in
Ref. \onlinecite{tserkovnyak-2006} (Eq.~(9) there) for a
particular case of a collinear AFM. We stress that this effect, in
contrast to those considered above, reflects the action of AFM (as
potential polarizer) on the non-polarized current.

Current-induced STT in AFMs can be used for the description of
current-induced dynamics of the domain walls and other
inhomogeneous states in the conducting AFMs and other materials
with the nontrivial magnetic structure. It should be noted that
thus introduced dissipative torque accounts for the rotation (in
space or in time) of the magnetic sublattices only and does not
take into account flip processes (current-induced transition from
AFM to FM structure) that could take place at the substantially
higher current density.

The last question that is worth to discuss here is the effect of
the ac current that results from the direct $sd$-exchange between
free and localized electrons (the effective magnetic field
$\mathbf{H}^{\mathrm{curr}}_{\mathrm{eff}}$,
Eq.~(\ref{eq_current-field})). As it follows from the dynamic
Eqs.~(\ref{eq_noncollinear_frequency}),
(\ref{eq_parallel_rotation}), \emph{time-derivative} $dj/dt$ of
spin-polarized current works as a driving force for oscillations
of AFM moments. Its value growth linearly with the current
frequency. This effect has no counterpart in FM, where the
analogous field-like torque is proportional to $j$, but not to
$dj/dt$. Thus, in AFM the current-induced movements of AFM vectors
are controlled by two ``forces'' which are 90$^\circ$ shifted in
phase. The phase shift between the current and system response
depends upon the frequency. This opens an additional way for
experimental observation of current-induced phenomena in AFMs.

E.V.G. is grateful B. Ivanov for valuable discussion.  The authors
acknowledge the financial support from the Physics and Astronomy
Department of
 National Academy of Science of Ukraine in the framework of the Special Program for
 Fundamental Researches. The work was partially supported by the
 grant from Ministry of Education and Science of Ukraine.

\appendix

\section{Description of macroscopic magnetic dynamics with the help of the Gibbs' vector}\label{appen_A}
The Gibbs' vector $\boldsymbol\varphi$ gives a convenient way for
parametrization of the rotational symmetry group $O$(3). Group
multiplication (symbol $\circ$) of two rotations,
$\boldsymbol\varphi_1$ and than $\boldsymbol\varphi_2$, is
noncommutative and is given by the rule:
\begin{equation}\label{eq_multiplication_rotation}
\boldsymbol\varphi_2\circ\boldsymbol\varphi_1\equiv\frac{\boldsymbol\varphi_2+\boldsymbol\varphi_1+[\boldsymbol\varphi_2\times\boldsymbol\varphi_1]}{1-\boldsymbol\varphi_2\boldsymbol\varphi_1}.
\end{equation}

In application to the description of macroscopic magnetic dynamics
the vector $\boldsymbol\varphi$ is a field variable that
characterizes the state of magnetic system at a time moment $t$ and
in a space point  $\mathbf{r}$ respective to some reference
(equilibrium) state. ``Radius-vector'' $\delta\boldsymbol\theta/2$
between two states at different moments, $\boldsymbol\varphi(t)$ and
$\boldsymbol\varphi(t+dt)=\boldsymbol\varphi(t)+d\boldsymbol\varphi$
is given by Eq.~(\ref{eq_rotation_angle_1})  and can be found from
Eq.~(\ref{eq_multiplication_rotation}) if we set
\begin{equation}\label{eq_distance_rotation}
\frac{\delta\boldsymbol\theta}{2}=[\boldsymbol\varphi(t)+d\boldsymbol\varphi]\circ[-\boldsymbol\varphi(t)].
\end{equation}

``Radius-vector'' between two states in different points
$\mathbf{r}$ and $\mathbf{r}+d\mathbf{r}$ is found in a similar way.

Equilibrium magnetic structure of any AFM with strong exchange
coupling can be described with the use of up to three orthogonal
vectors, $\mathbf{L}^{(0)}_k$ ($k$=1,2,3). Once $\mathbf{L}^{(0)}_k$
 are known, orientation of AFM vectors at any $t$ and $\mathbf{r}$, $\mathbf{L}_k(t,\mathbf{r})$, can be expressed
in terms of the Gibbs's vector as follows:
\begin{equation}\label{eq_rotation_vector}
\mathbf{L}_k=\mathbf{L}^{(0)}_k+\frac{2}{1+\boldsymbol\varphi^2}\left[[\boldsymbol\varphi\times[\boldsymbol\varphi\times\mathbf{L}^{(0)}_k]]+[\boldsymbol\varphi\times\mathbf{L}^{(0)}_k]\right].
\end{equation}

The potential energy $U_{\mathrm{AFM}}(\boldsymbol\varphi)$ in
Eq.~(\ref{eq_Lagrange_1}) can be found in a following way. One
must construct a symmetry-invariant expression for
$U_{\mathrm{AFM}}(\boldsymbol\varphi)$ in terms the components of
$\mathbf{L}_k$ vectors and then express $\mathbf{L}_k$ in terms of
$\boldsymbol\varphi$ using the relation
(\ref{eq_rotation_vector}).

%

\end{document}